Element Substitution Effect on Superconductivity in $BiS_2$-Based $NdO_{1-x}F_xBiS_2$

Takafumi Hiroi, Joe Kajitani, Atsushi Omachi, Osuke Miura, and Yoshikazu Mizuguchi*

Department of Electrical and Electronic Engineering, Tokyo Metropolitan University, 1-1, Minami-osawa, Hachioji, 192-0397, Japan

Corresponding author: Y. Mizuguchi

E-mail: mizugu@tmu.ac.jp

Telephone: +81-42-677-2748

FAX: +81-42-677-2756

Abstract

  Recently, new layered superconductors having a $BiS_2$-type conduction layer have been discovered. $NdO_{1-x}F_xBiS_2$ is a typical $BiS_2$-based superconductor with a maximum $T_c$ of 5.4 K. In this study, the effect of element substitution within the superconducting layer of $BiS_2$-based $NdO_{0.5}F_{0.5}BiS_2$ was investigated. We systematically synthesized two kinds of polycrystalline samples of $NdO_{0.5}F_{0.5}Bi(S_{1-x}Se_x)_2$ and $NdO_{0.5}F_{0.5}Bi_{1-y}Sb_yS_2$ by a two-step solid-state reaction method. The phase purity and the changes in lattice constants were investigated by x-ray diffraction. The superconducting properties were investigated by magnetic susceptibility and electrical resistivity measurements. It was found that the partial substitution of S by Se resulted in the uniaxial lattice expansion along the $a$ axis. The superconducting transition temperature were gradually degraded with increasing Se concentration in NdO$_{0.5}$F$_{0.5}$Bi(S$_{1-x}$Se$_x$)$_2$. When Bi was partially substituted by Sb, the uniaxial lattice contraction along the $c$ axis was observed in NdO$_{0.5}$F$_{0.5}$Bi$_{1-y}$Sb$_y$S$_2$. In the Sb-substituted system, a metal-insulator transition was observed with increasing Sb concentration.

1. Introduction

Since the discovery of the novel layered superconductor $Bi_4O_4S_3$, many studies for exploring new superconductors and clarifying the mechanisms of superconductivity in the $BiS_2$ family have been carried out [1]. So far, three types of $BiS_2$-based superconductors, $Bi_4O_4S_3$ [1], $REOBiS_2$ (RE = La, Ce, Pr, Nd, Yb) [2-9], $SrFBiS_2$ [10,11], have been discovered. The highest record of the transition temperature ($T_c$) is 11 K in $LaO_{0.5}F_{0.5}BiS_2$ synthesized using a high-pressure technique. These $BiS_2$-based superconductors basically have a layered structure composed of common $BiS_2$ superconducting layers and the blocking layers, which is quite similar to those of the cuprate and the Fe-based superconductors [13,14]. The superconducting properties of the $BiS_2$ family can be tuned by changing the structure of blocking layer. To induce superconductivity in the $BiS_2$-based layered materials, electron doping into the $BiS_2$ layers is required [15-20]. For example, in the $LaOBiS_2$ system, superconductivity is induced by a partial substitution of $O^{2-}$ by $F^-$ or a partial substitution of $La^{3+}$ by $M^{4+}$ [2,21] at the blocking layers. Namely, the element substitution at the blocking layer is effective to induce/tune superconductivity in the $BiS_2$-based family. Furthermore, it was reported that the superconducting properties are sensitive to the change in the local crystal structures [22]. For the $REOBiS_2$ systems (RE = La, Ce and Pr), the optimal (the highest) superconducting properties are obtained by optimizing the crystal structures by using a high-pressure technique [2,4,6,23-27].

Although there are many reports on element substitution effects of blocking layers in $REO_{1-x}F_xBiS_2$, the effect of element substitution at the superconducting layer has been reported in only one report. In $Bi_4O_4S_3$, partial substitution of S by Se decreased its $T_c$ [28]. In this study, the effect of element substitution at the superconducting layers on superconductivity of $NdO_{0.5}F_{0.5}BiS_2$ was investigated. We synthesized Se-substituted $NdO_{0.5}F_{0.5}Bi(S_{1-x}Se_x)_2$ and Sb-substituted $NdO_{0.5}F_{0.5}Bi_{1-y}Sb_yS_2$. These element substitutions basically generate the changes in the lattice constants without any change in the carrier concentration.

2. Experimental

In this study, the F concentration was fixed to be $x = 0.5$ because the highest $T_c$ was obtained for $x = 0.5$ in $NdO_{1-x}F_xBiS_2$. The polycrystalline samples of $NdO_{0.5}F_{0.5}Bi(S_{1-x}Se_x)_2$ and $NdO_{0.5}F_{0.5}Bi_{1-y}Sb_yS_2$ ($x$, $y = 0$, 0.1 and 0.2) were prepared by the two-step solid-state reaction method. The starting materials for $NdO_{0.5}F_{0.5}Bi(S_{1-x}Se_x)_2$ are $Nd_2O_3$ powders, $Nd_2S_3$ powders, $NdF_3$ powders, Bi grains, $BiSe_2$ powders and S grains. The starting materials for $NdO_{0.5}F_{0.5}Bi_{1-y}Sb_yS_2$ are $Nd_2O_3$ powders, $Nd_2S_3$ powders, $NdF_3$ powders, Bi grains, Sb powders and S grains. The mixtures with the nominal compositions of $NdO_{0.5}F_{0.5}Bi(S_{1-x}Se_x)_2$ and $NdO_{0.5}F_{0.5}Bi_{1-y}Sb_yS_2$ were mixed-well, ground, pelletized, sealed in an evacuated quartz tube and heated at 800 °C for 15 h. The obtained products were ground, pelletized, sealed in an evacuated quartz tube and annealed at 800 °C for 15 h again. The obtained samples were characterized by powder X-ray diffraction (XRD) using the $\theta$ - $2\theta$ method with a CuKα radiation. The lattice constants were calculated using the peak positions. The temperature dependence of magnetization was measured by a superconducting quantum interface device (SQUID) magnetometer with an applied field of 5 Oe. The temperature dependence of electrical resistivity was measured by the four-probe method.

3. Results

Figure 1 shows the powder XRD patterns of (a)NdO$_{0.5}$F$_{0.5}$Bi(S$_{1-x}$Se$_x$)$_2$ and (b)NdO$_{0.5}$F$_{0.5}$Bi$_{1-y}$Sb$_y$S$_2$ ($x$, $y$ = 0, 0.1 and 0.2). These XRD patterns were indexed using the *P*4/*nmm* space group. In the XRD patterns of NdO$_{0.5}$F$_{0.5}$Bi(S$_{1-x}$Se$_x$)$_2$, impurity peaks appeared in the doping range of $y \geq 0.1$. In contrast, the impurity peaks were not detected upon Sb substitution in the XRD patterns of NdO$_{0.5}$F$_{0.5}$Bi$_{1-y}$Sb$_y$S$_2$. Almost all of the peaks became slightly broader in the substituted samples. The lattice constants *a* and *c* were calculated using the positions of the (200) peaks and the (004) peaks, respectively.

The Se-concentration dependences of the lattice constants of the *a* axis and *c* axis for NdO$_{0.5}$F$_{0.5}$Bi(S$_{1-x}$Se$_x$)$_2$ is shown in Figs. 2(a) and (b), respectively. With increasing Se concentration, the lattice constant of the *a* axis increased while the lattice constants of the *c* axis did not change obviously in NdO$_{0.5}$F$_{0.5}$Bi(S$_{1-x}$Se$_x$)$_2$. The Se substitution for the S site resulted in a uniaxial lattice expansion along the *a* axis.

The Sb-concentration dependences of the lattice constants of the *a* axis and *c* axis for NdO$_{0.5}$F$_{0.5}$Bi$_{1-y}$Sb$_y$S$_2$ is shown in Figs. 2(a) and (b), respectively. With increasing Sb concentration, the lattice constants of the *a* axis did not change obviously while the lattice constant of the *c* axis decreased in NdO$_{0.5}$F$_{0.5}$Bi$_{1-y}$Sb$_y$S$_2$. The Sb substitution for the Bi site resulted in a uniaxial lattice contraction along the *c* axis.

The temperature dependences of magnetic susceptibility ($\chi$) of NdO$_{0.5}$F$_{0.5}$Bi(S$_{1-x}$Se$_x$)$_2$ ($x$ = 0, 0.1 and 0.2) are shown in Fig. 3. The inset shows the Se-concentration dependences of the $T_{c\_onset}$ and the $T_{c\_10\%}$. The $T_{c\_10\%}$ is defined as a temperature where the value of $\chi$ becomes 10 % of $\chi$(2 K). The $T_{c\_onset}$ and $T_{c\_10\%}$ decreased with increasing Se concentration. The shielding volume fraction of NdO$_{0.5}$F$_{0.5}$Bi(S$_{1-x}$Se$_x$)$_2$ also decreased drastically. On the basis of these results, we concluded that the Se substitution for the NdO$_{0.5}$F$_{0.5}$BiS$_2$ superconductor degraded superconducting properties and finally destroyed the superconducting states.

For the NdO$_{0.5}$F$_{0.5}$Bi$_{1-y}$Sb$_y$S$_2$ system, superconductivity was strongly suppressed by the Sb substitution. Even for $y$ = 0.1, superconducting transition was not observed in the magnetic susceptibility measurements. Therefore, we performed electrical resistivity measurements to investigate the Sb substitution effect on the physical properties. Figure 4 shows the temperature dependences of electrical resistivity for NdO$_{0.5}$F$_{0.5}$Bi$_{1-y}$Sb$_y$S$_2$. The resistivity drastically increased with increasing Sb concentration. When the Sb concentration was $y$ = 0.1, the value of resistivity was $1.90\times10^2$ m$\Omega$cm at 6 K. When the Sb concentration was $y$ = 0.2, resistivity was $1.35\times10^5$ m$\Omega$cm at 6 K. The value of resistivity for $y$ = 0.2 was about $10^3$ times larger than that of $y$ = 0.1. The remarkable change suggests that the Sb substitution for the Bi site results in the metal-insulator transition in the NdO$_{1-x}$F$_x$BiS$_2$ system. Having considered the one-dimensional character of the electronic states in the BiS$_2$-based materials [15], we speculate that the observed metal-insulator transition would be related to the Anderson localization. To investigate further, detailed studies on transport properties and crystal structure using single crystals are needed.

4. Conclusions

The effect of the element substitution within the superconducting layer on superconductivity in the BiS$_2$-based superconductor NdO$_{0.5}$F$_{0.5}$BiS$_2$ was investigated. The polycrystalline samples of NdO$_{0.5}$F$_{0.5}$Bi(S$_{1-x}$Se$_x$)$_2$ and NdO$_{0.5}$F$_{0.5}$Bi$_{1-y}$Sb$_y$S$_2$ ($x$, $y$ = 0, 0.1 and 0.2) were synthesized by the two-step solid state reaction method. When S was substituted by Se, the lattice constant of the $a$ axis increased, and the superconducting properties ($T_c$ and shielding volume fraction) were degraded. When Bi was substituted by Sb, the lattice constant of the $c$ axis decreased, and a metal-insulator transition was observed. The element substitution within the superconducting layer degrades superconductivity in the NdO$_{1-x}$F$_x$BiS$_2$ system.

Acknowledgements

The authors would like to thank Prof. K. Kuroki of Osaka University for fruitful discussion. This work was partly supported by JSPS KAKENHI Grant Numbers 25707031, 26600077.

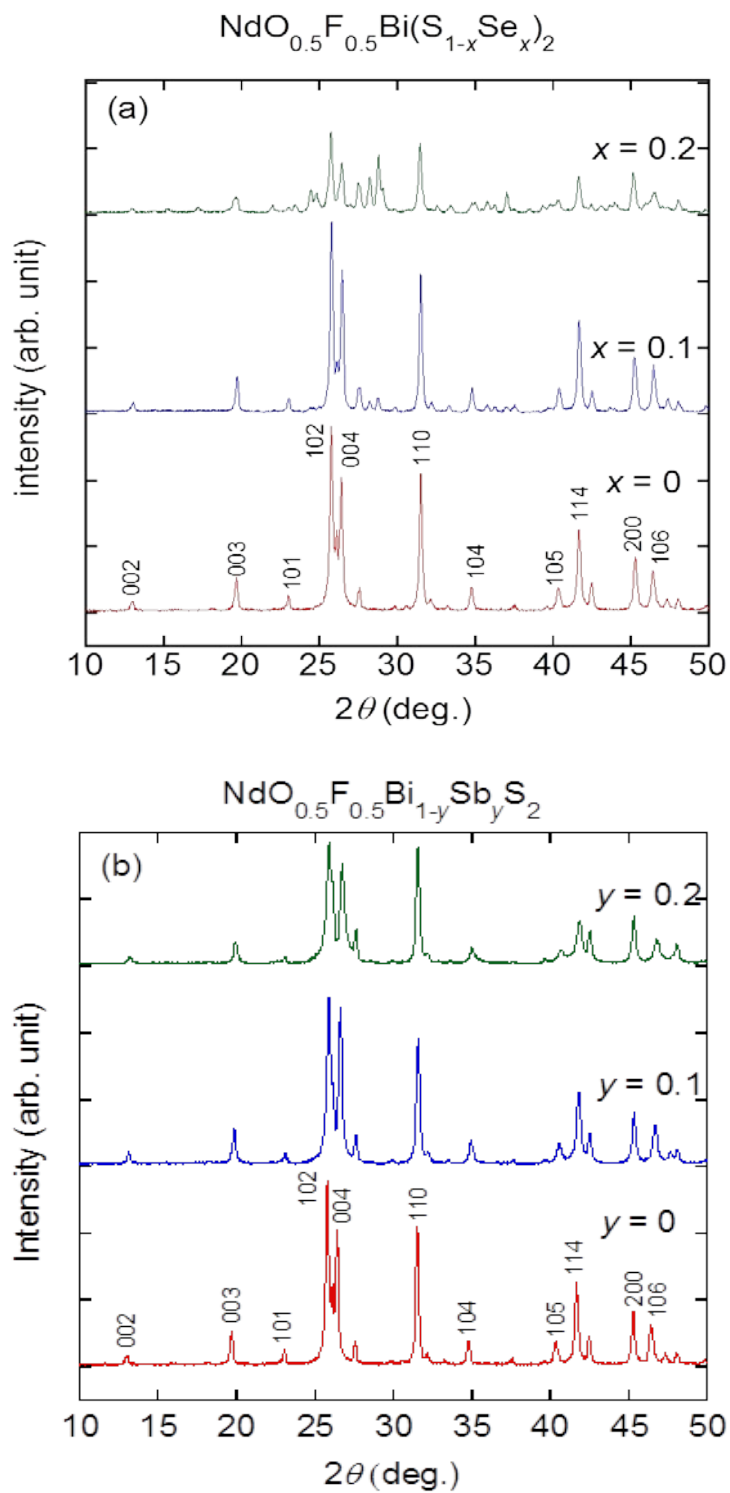

Fig. 1 XRD patterns of (a) $NdO_{0.5}F_{0.5}Bi(S_{1-x}Se_x)_2$ ($x$ = 0, 0.1, 0.2) and (b) $NdO_{0.5}F_{0.5}Bi_{1-y}Sb_yS_2$ ($y$ = 0, 0.1, 0.2).

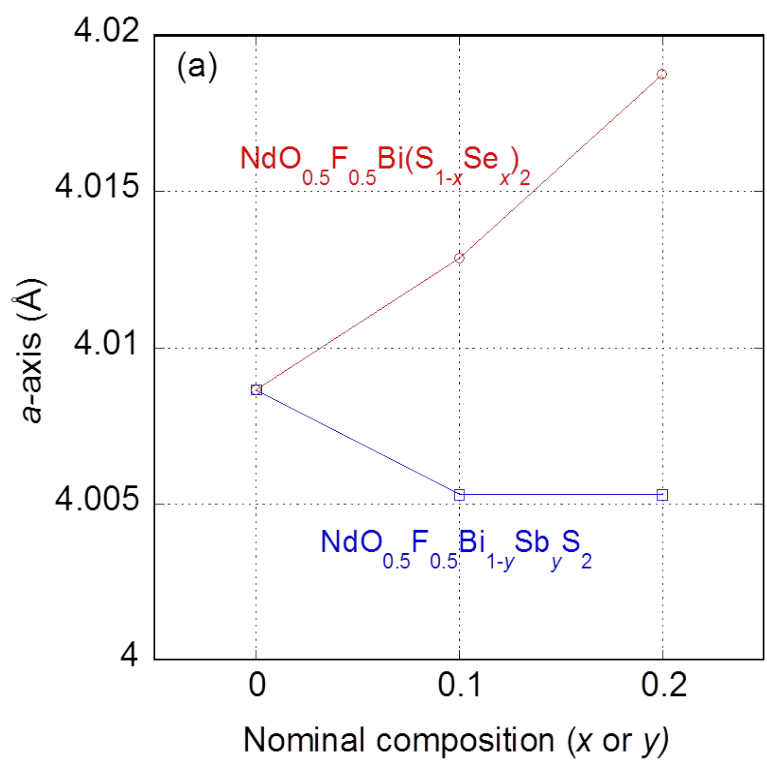

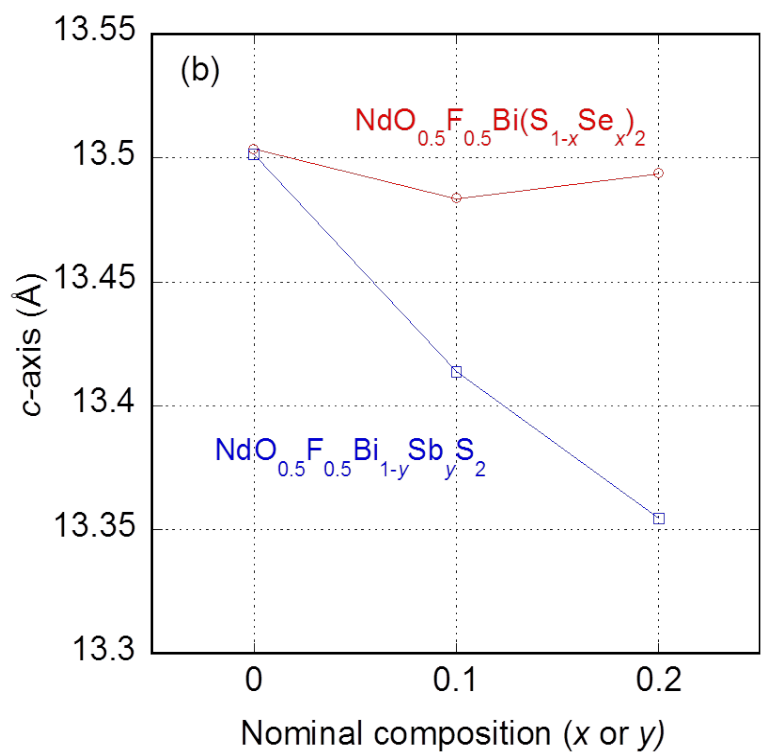

Fig. 2 Se-concentration $x$ dependence (or Sb-concentration $y$ dependence) of the lattice constants of (a) $a$-axis and (b) $c$-axis.

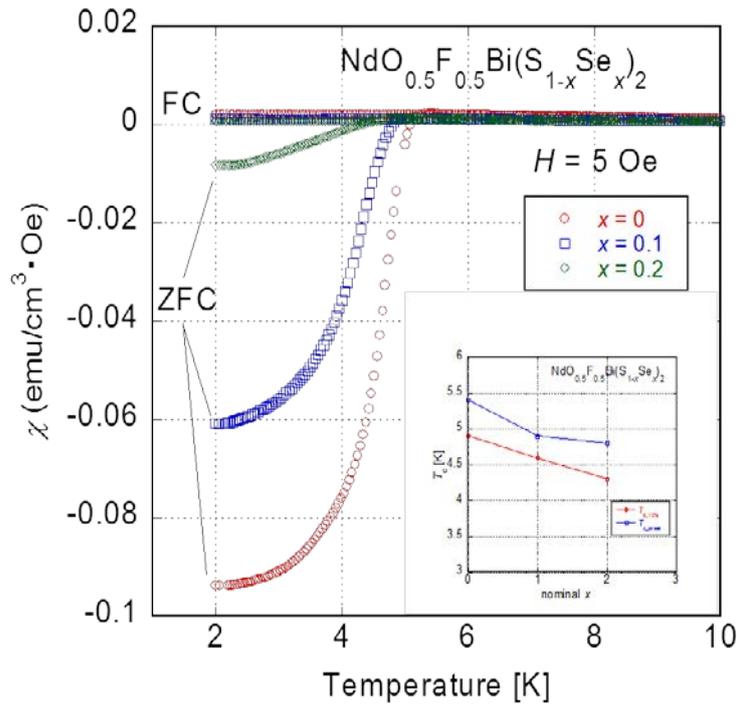

Fig. 3 Temperature dependences of the magnetic susceptibility ($\chi$) of $NdO_{0.5}F_{0.5}Bi(S_{1-x}Se_x)_2$ ($x = 0, 0.1, 0.2$).

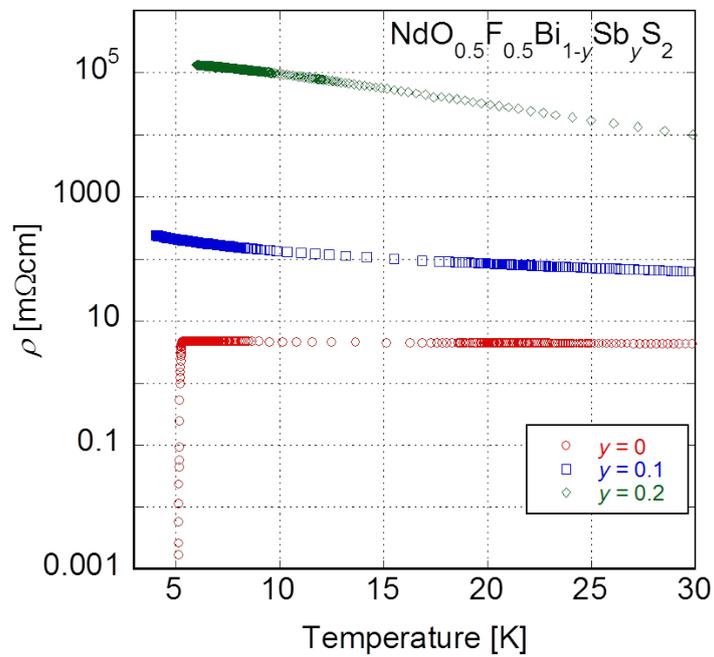

Fig. 4 Temperature dependences of electrical resistivity ($\rho$) of $NdO_{0.5}F_{0.5}Bi_{1-y}Sb_yS_2$ ($y = 0, 0.1, 0.2$).